\documentclass[twocolumn,showpacs,preprintnumbers,amsmath,amssymb]{revtex4}

\usepackage{graphicx}
\usepackage{dcolumn}
\usepackage{bm}

\begin{document}

\setlength{\topmargin}{20mm}
\addtolength{\topmargin}{-1in}

\title{ Coherent vs Thermally Activated Singlet Exciton Fission in Acene Derivatives From First Principles Quantum Dynamics Simulations: Molecular Packing Makes the Difference }

\author{Hiroyuki Tamura$^{1,*}$}
\author{Miquel Huix-Rotllant$^2$}
\author{Irene Burghardt$^2$}
\author{Yoann Olivier$^3$}
\author{David Beljonne$^3$}

\affiliation{$^1$WPI-Advanced Institute for Material Research, 
Tohoku University, 2-1-1 Katahira, Aoba-ku, Sendai, 980-8577, Japan}

\affiliation{$^2$Institute of Physical and Theoretical Chemistry, Goethe University Frankfurt, Max-von-Laue-Str. 7, 60438 Frankfurt/Main, Germany}

\affiliation{$^3$Laboratory for Chemistry of Novel Materials, University of Mons, Place du Parc 20, 7000 Mons, Belgium}

\date{\today}
\begin{abstract}
The mechanisms underlying coherent and thermally activated singlet exciton fission in $\pi$-stacked acene crystals are clarified based on quantum dynamics simulations parameterized against a highly correlated description of the electronic excitations and their couplings to intramolecular and intermolecular vibrations. In TIPS-pentacene crystals, the relative longitudinal shift of the molecular backbones yields large electronic couplings of the triplet exciton pair with both the singlet exciton and charge transfer (CT) states. CT-mediated superexchange and direct pathways are found to contribute synergetically to the ultrafast ($\sim$100fs) singlet fission process driven by vibronic coherences. By contrast, the electronic couplings for singlet fission strictly vanish at the equilibrium $\pi$-stacking of rubrene that exhibits $C_{2h}$ symmetry. In this case, the process is incoherent and driven by excitations of symmetry-breaking intermolecular vibrations, rationalizing the experimentally observed temperature dependence.
 
\end{abstract}


\maketitle

Singlet excitons in certain molecular crystals such as acenes and their derivatives can split into two triplet excitons following a spin-conserving process known as singlet fission (SF) \cite{ChemRev, RubreneNatMater, RubrenePRL, Science, Pentacene, RubrenePRB, Zimmerman, Mons, Yost, Reichman, Thoss, TIPS2}. SF has attracted a lot of attention lately in the context of organic photovoltaics, since it permits internal quantum efficiency in excess of 100$\%$ through the conversion of high-energy photons into two excitons and subsequently two electron-hole pairs at donor-acceptor interfaces \cite{Science}. As triplet excitons usually diffuse over longer distances ($\sim$$\mu$m) compared to singlets \cite{RubreneNatMater, RubrenePRL}, SF can also be implemented in thick multilayer architectures.

Transient absorption spectroscopy investigations point to an ultrafast SF process in molecular crystals of pentacene \cite{Pentacene} and TIPS-pentacene \cite{TIPS2}. Contradicting views on the mechanistic aspects of SF in these crystals have been reported in the literature \cite{Zimmerman, Mons, Yost}. Briefly, it has been proposed that SF either proceeds through a direct two-electron coupling between the singlet exciton (XT) and the triplet pair (TT) or follows an indirect mechanism where charge-transfer (CT) states act as virtual mediating states (i.e. superexchange) or are transiently populated. In both cases, the singlet-to-triplet conversion is believed to be driven by coupling to nuclear degrees of freedom and involves either a conical intersection or an avoided crossing pathway. Very interestingly, recent two-dimensional photon echo experiments performed on TIPS-pentacene thin films have demonstrated that: (i) while light absorption does not directly generate a coherent superposition of XT and TT at time zero, the TT population grows in time at the expense of XT with an 80fs time constant; and that (ii) yet, the vibrational coherences generated upon photoexcitation into XT are largely transferred to TT during the SF process \cite{TIPS2}. This implies an essential role of electron-phonon (vibronic) coupling in the coherent SF process. In contrast to TIPS-pentacene, SF in the rubrene crystal has been reported to be a thermally activated process and is characterized by a much longer time constant of few ps \cite{RubrenePRB}. The origin of the different SF mechanisms in these two families of molecular crystals is still unclear. 

In this study, we clarify the SF mechanisms in TIPS-pentacene and rubrene based on non-adiabatic quantum dynamics simulations fully parameterized against highly correlated {\it ab initio} electronic structure calculations. We find that while ultrafast SF in TIPS-pentacene occurs mostly via a CT-mediated mechanism that preserves vibronic coherences, the corresponding process in rubrene is incoherent and thermally activated by intermolecular symmetry-breaking vibrations (phonons).

\begin{figure}[!ht]
\includegraphics[width=6.7cm]{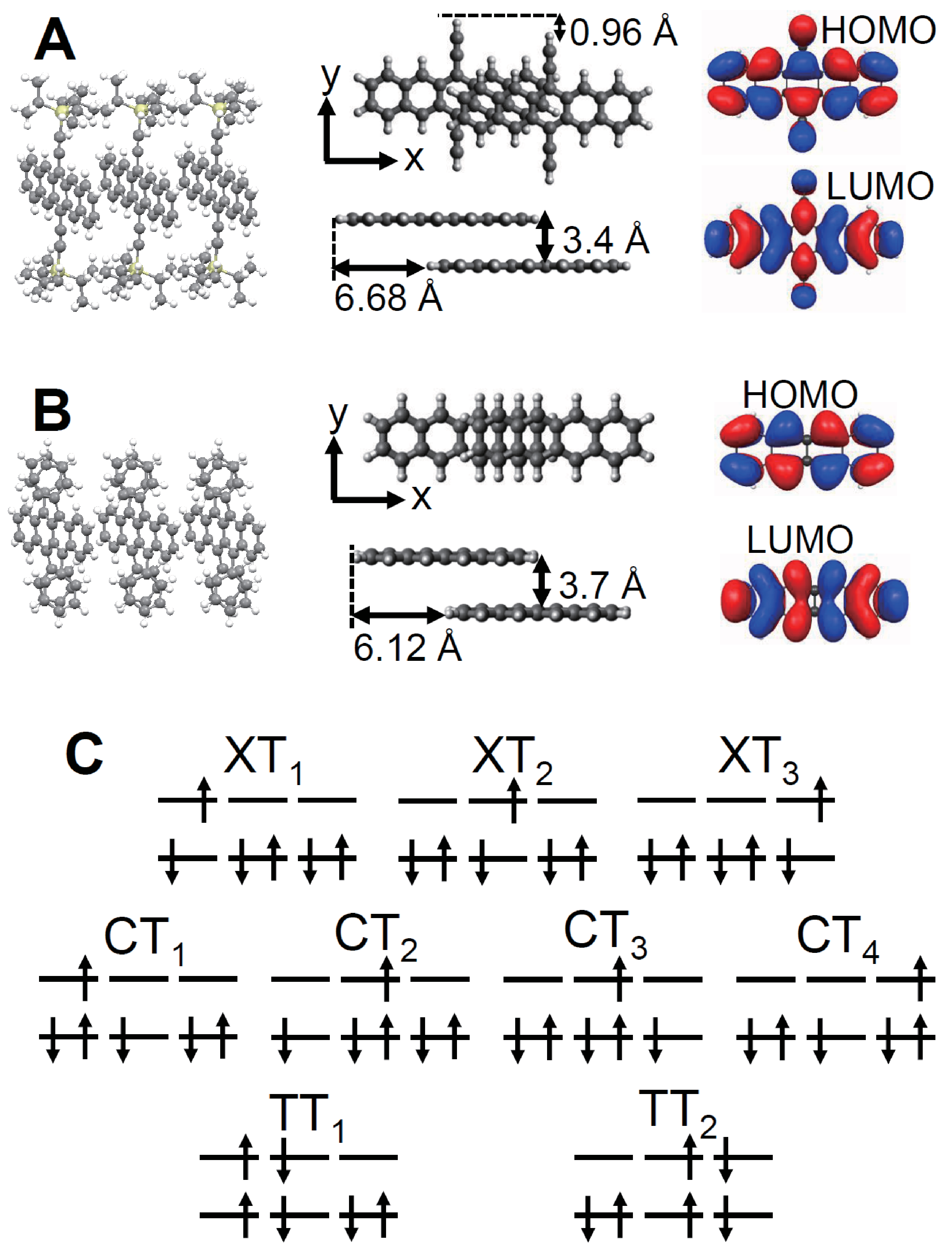}
\caption{
\label{fig1} 
(Color online)
(A) Crystal structure of TIPS-pentacene and the dimer model for MRMP2 calculations, where the side group is simplified as alkyne. The active orbitals of dimer are composed of the bonding and anti-bonding hybridizations of the HOMO and LUMO of two monomers. (B) Crystal structure of rubrene and the dimer model for MRMP2 calculations, where the side group is neglected. (C) Diagram of electron configurations of XT, TT, and CT states in the 3-site, 9-state model. 
}
\end{figure}

We consider the following linear vibronic coupling Hamiltonian in a diabatic representation:

\begin{eqnarray}
  \label{H}
\mbox{} &  & \hspace*{0.0cm}
H = \sum_I h_I({\bf x}) | I \rangle \langle I |
+ \sum_{I>J} V_{IJ}(X) ( | I \rangle \langle J | + h.c. )
\end{eqnarray}
\begin{eqnarray}
  \label{h}
\mbox{} &  & \hspace*{0.0cm}
h_I({\bf x}) = \sum_i \frac{\omega_i}{2} (x_i^2 + p_i^2)
+ \sum_i \kappa_i^I x_i + E_I
\end{eqnarray}

\noindent where $I$ and $J$ are the indices for the relevant XT, TT, and CT states. The diagonal term, $h_I({\bf x})$, consists of harmonic oscillators for the intra-molecular modes, $x_i$, where $\omega_i$ is the frequency, $p_i$ is the momentum, $\kappa_i^I$ is the vibronic coupling, $E_I$ is the excitation energy at the ground state geometry. The off-diagonal term, $V_{IJ}$, is the electronic coupling between the states which depends on inter-molecular modes, $X$. The multi-configuration time-dependent Hartree (MCTDH) method \cite{MCTDH} has been used to solve the Schr\"{o}dinger equation, 
$ i \partial \bm{\Psi} / \partial t = H \bm{\Psi} $, with 
$ \bm{\Psi} = \sum_I \Psi_I | I \rangle $, where $\Psi_I$ are the vibrational wavefunctions on the respective electronic states. 
All intra-molecular vibrational modes and selected inter-molecular modes are included in the MCTDH calculations. Our model can describe coherent and incoherent SF dynamics mediated by vibronic coupling of realistic systems.

We consider a molecular trimer, i.e., 3-site, 9-state model consisting of 3 XT, 2 TT, and 4 CT states (Fig. 1C) for the quantum dynamics calculations, where the Frenkel exciton on the center site (XT$_2$) is coupled to 2 TT states similar to the bulk condition. The Hamiltonian matrix elements (Table 1) are determined based on electronic structure calculations of the dimer models (Fig. 1A and 1B) using the multi-reference second order perturbation theory (MRMP2) with the correlation consistent polarized basis set (cc-pVDZ), expected to provide an accurate description of the excited states. The XT, TT and CT energies and their pairwise couplings are obtained by applying a unitary transformation from the dimer adiabatic electronic states to the diabatic representation \cite{SM}. The four frontier molecular orbitals, i.e. highest occupied molecular orbital (HOMO), HOMO-1, lowest unoccupied molecular orbital (LUMO) and LUMO+1, are considered as active orbitals in MRMP2 calculations on the dimers. The intra-molecular vibronic couplings are evaluated based on the frequencies of normal modes and the geometry optimizations in the respective states, using the density functional theory (DFT) with the PBE functional, where the spectral density is broaden considering the frequency splitting \cite{SM}. The GAMESS code \cite{GAMESS} is used for all {\it ab initio} calculations. 

First, we address SF in TIPS-pentacene. Table 1 shows the diabatic state energies and electronic couplings, as obtained from MRMP2 calculations in the ground-state geometry. SF in TIPS-pentacene is clearly exothermic. Accounting for the intra-molecular reorganization energy from the ground state to the exciton ($\lambda_{XT}$ = 0.15 eV) and TT ($\lambda_{TT}$ = 0.45 eV) state geometries, the driving force $\Delta$$E_{XT-TT}$ amounts to -0.45 eV. 

\begin{table}[h]
\caption{Hamiltonian matrix (symmetric matrix) of SF in diabatic representation where the diagonal and off-diagonal elements correspond to the excitation energies and the electronic couplings (eV), respectively.}

\begin{tabular}{cccccccccc} \hline \hline
& TT$_1$ & TT$_2$ & XT$_1$ & XT$_2$ & XT$_3$ 
& CT$_1$ & CT$_2$ & CT$_3$ & CT$_4$ \\ \hline 
TT$_1$ & E$_{TT}$ & 0.0 & -V$_{TX}$ & V$_{TX}$ & 0.0
& V$_{TC}$ & V$_{TC}$ & 0.0 & 0.0 \\
TT$_2$ & & E$_{TT}$ & 0.0 & V$_{TX}$ & -V$_{TX}$ 
& 0.0 & 0.0 &V$_{TC}$ & V$_{TC}$ \\
XT$_1$ & & & E$_{XT}$ & V$_{XX}$ & 0.0 
& -V$_{H}$ & V$_{L}$ & 0.0 & 0.0 \\
XT$_2$ & & & & E$_{XT}$ & V$_{XX}$  
& -V$_{L}$ & V$_{H}$ & V$_{H}$ & -V$_{L}$ \\
XT$_3$ & & & & & E$_{XT}$ & 0.0 & 0.0 & V$_{L}$ & -V$_{H}$  \\
CT$_1$ & & & & & & E$_{CT}$ & -V$_{CC}$ & -V$_{CC}$ & 0.0  \\
CT$_2$ & & & & & & & E$_{CT}$ & 0.0 & -V$_{CC}$  \\
CT$_3$ & & & & & & & & E$_{CT}$ & -V$_{CC}$  \\
CT$_4$ & & & & & & & & & E$_{CT}$ \\ \hline \hline
\end{tabular}

\begin{tabular}{cccc} \hline \hline
& E$_{TT}$ & E$_{XT}$ & E$_{CT}$ \\ \hline 
TIPS-Pentacene & 1.469 & 1.618 & 1.992 \\
Rubrene      & 2.451 & 2.311 & 3.097 \\ \hline \hline
\end{tabular}

\begin{tabular}{ccccccc} \hline \hline
& V$_{TX}$ & V$_{TC}$ & V$_{XX}$ & V$_{H}$ & V$_{L}$ 
& V$_{CC}$  \\ \hline 
TIPS-Pentacene 
& 0.013 & 0.084 & 0.018 & 0.041 & 0.149 & 0.017 \\
Rubrene 
& 0.0 & 0.0 & 0.079 & -0.175 & 0.086 & 0.035 \\ \hline \hline
\end{tabular}

\end{table}

The slipped-stacked configuration of TIPS-pentacene results in large electronic couplings for both the direct and CT-mediated pathways (Table 1). Accordingly, the quantum dynamics simulations point to an ultrafast SF with a timescale of $\sim$100 fs (Fig. 2), in very good agreement with experimental data \cite{TIPS2}. The contributions arising from the two channels can be clearly disentangled by accounting for only one of either the direct two-electron XT-TT coupling or the indirect XT-CT and CT-TT couplings (Fig. 2B and 2C). Even though the intermediate CT states are $\sim$0.4 eV higher in energy than the initial exciton state in our calculations, the superexchange pathway is found to promote an ultrafast SF process and dominates over the direct mechanism, at least in the first 100fs. Such a coherent superexchange cannot be described by perturbative hopping pictures, highlighting the importance of including electron-electron and electron-vibration couplings on an equal footing when solving the time evolution of the system. Note that the picture above is not affected by initial excitation conditions, i.e. very similar results are obtained when preparing the system either in a localized Frenkel exciton or in as a delocalized (bright or dark) exciton state (Fig. 2A, 2D, and 2E). Besides, the ultrafast SF in TIPS-pentacene is robust against $\Delta$$E_{XT-TT}$ as long as the system is exothermic \cite{SM}. 

\begin{figure}[!ht]
\includegraphics[width=7.5cm]{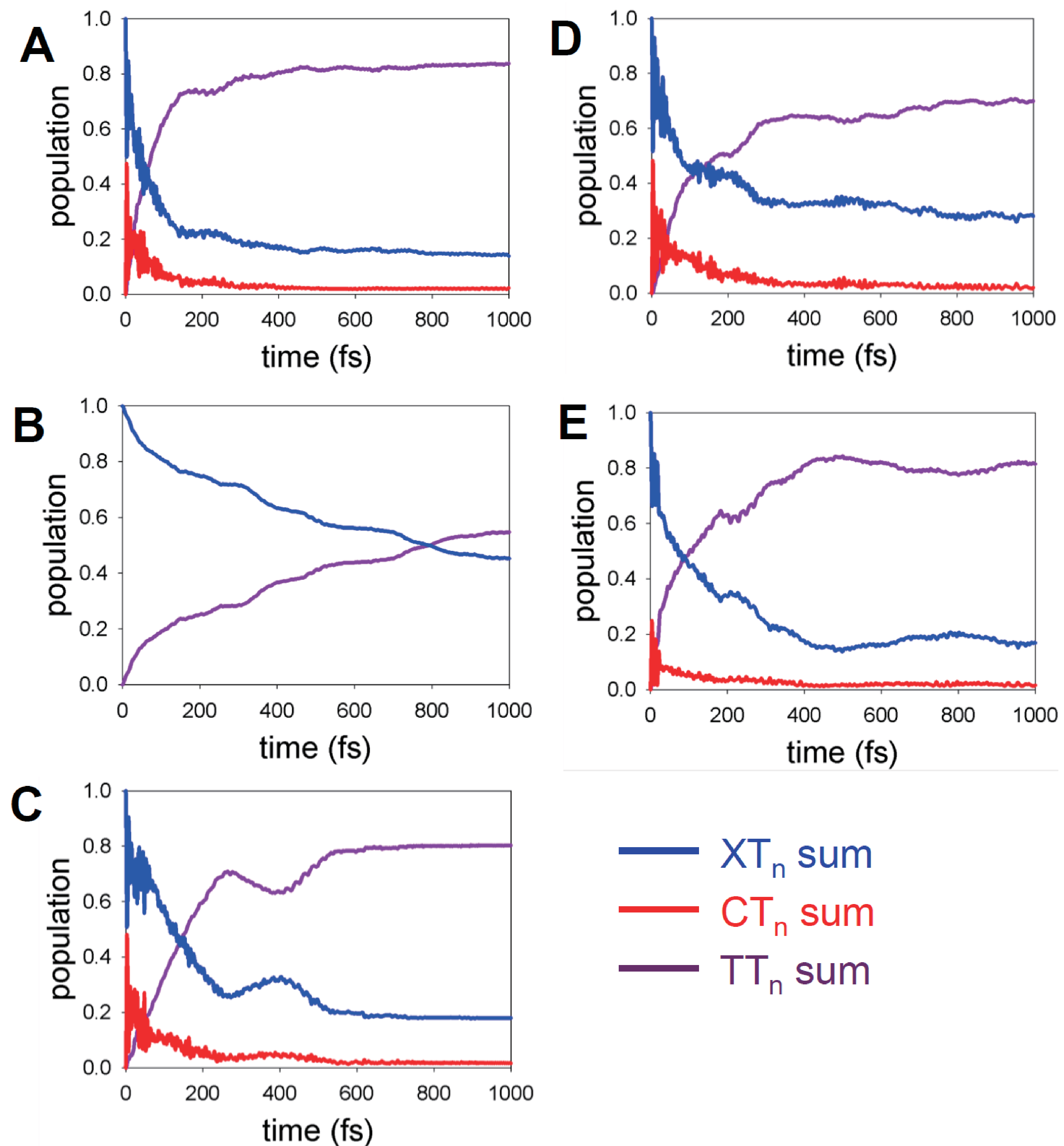}
\caption{
\label{fig2} 
(Color online)
Cumulative populations of XT$_n$ (blue), CT$_n$ (red), and TT$_n$ (purple) in the quantum dynamics calculations of SF in TIPS-pentacene. (A) Initial exciton is localized on the center molecule (XT$_2$), and all the electronic couplings are considered. Only one of (B) direct and (C) CT-mediated pathways are considered. Dynamics from (D) bright delocalized exciton with the initial amplitude, (XT$_1$, XT$_2$, XT$_3$) = (1/$\sqrt{3}$, 1/$\sqrt{3}$, 1/$\sqrt{3}$), and from (E) dark exciton, (XT$_1$, XT$_2$, XT$_3$) = (1/$\sqrt{3}$, -1/$\sqrt{3}$, 1/$\sqrt{3}$).
}
\end{figure}

The coherent nature of SF in TIPS-pentacene is analyzed based on the overlap of vibronic wave packets on the XT and TT states, which provides a measure of the coherence \cite{Coherence}. The substantial vibronic coherence during the first $\sim$100 fs (Fig. 3A) indicates that the wave packet prepared on the XT potential energy surface is only weakly perturbed upon crossing onto the TT hypersurface, in line with the coherent nature of the process revealed in the photon echo experiments by Rao et al. \cite{TIPS2}. Here, the vibronic coupling mediates resonance between the initial and final states, and thus purely electronic models cannot correctly describe the SF process \cite{SM}. The spread of the wave packet on the XT potential surface is wide enough to cross the XT-TT avoided crossing immediately after the excitation from the ground state (Fig. 3B), hence rationalizing the coherence transfer via the direct pathway (Fig. 3A). Our analysis further clarified that the vibronic coherence is transferred throughout the superexchange pathway even via the higher-lying CT states (Fig. 3A) owing to the strong electronic couplings. 

\begin{figure}[!ht]
\includegraphics[width=7.5cm]{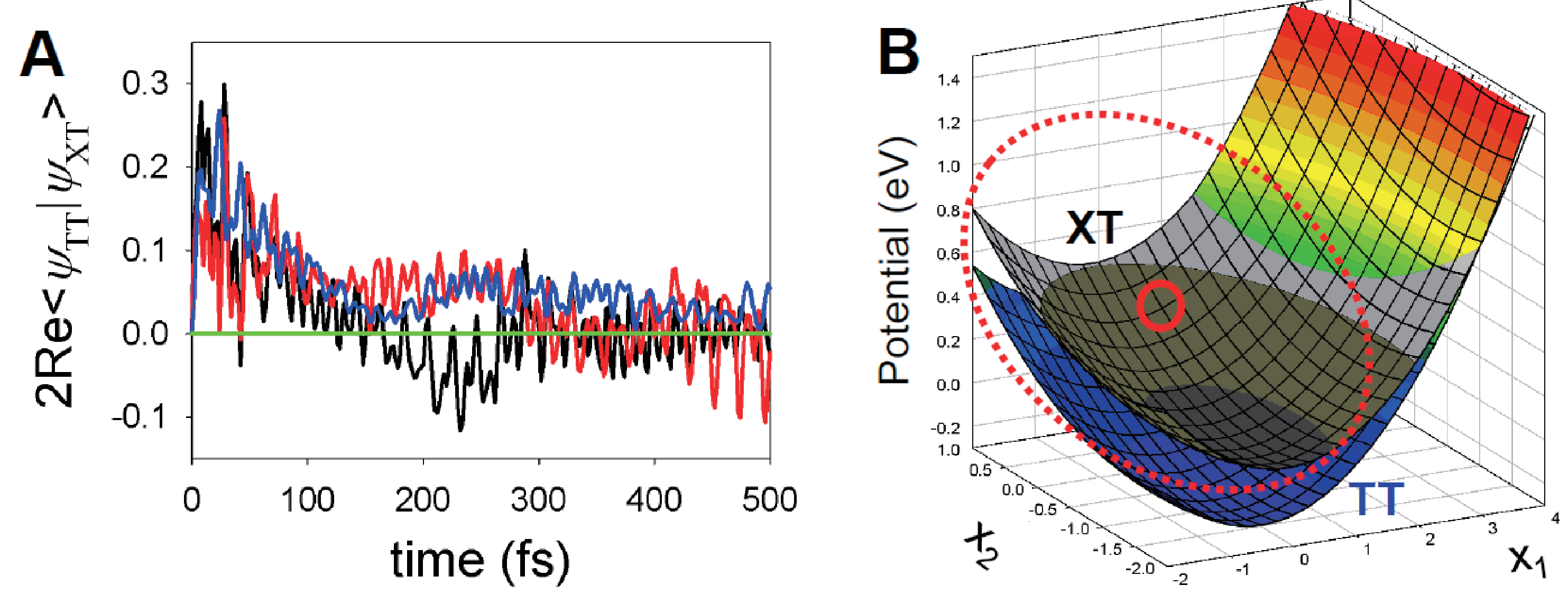}
\caption{
\label{fig3} 
(Color online)
(A) Overlap integral of vibrational wave packets on the XT and TT states during SF in TIPS-pentacene (black), and those considering only the direct (blue) and superexchange (red) pathways, where the initial exciton is localized on XT$_2$. 2Re$ \langle \Psi_{TT} | \Psi_{XT} \rangle $ for rubrene (green) is negligibly small. (B) Potential energy surface of TT (color) and XT (gray scale) states of TIPS-pentacene in the effective-modes representation \cite{PRL}. The solid and dashed red circles indicate the Frank-Condon region from the ground state and the spread of wave packet, respectively. 
}
\end{figure}

Next, we investigate the origin for the reported thermally activated SF process in rubrene crystals. SF in rubrene is exothermic ($\Delta$$E_{XT-TT}$ = -0.21 eV) when accounting for the reorganization energies ($\lambda_{XT}$ = 0.18 eV and $\lambda_{TT}$ = 0.53 eV), although the driving force is smaller than pentacene. Remarkably, the TT-XT and TT-CT couplings are strictly zero at the $C_{2h}$ equilibrium geometry (V$_{TX}$ and V$_{TC}$ in Table 1), hence SF in rubrene is expected to be activated by some distortion breaking the symmetry. We show below that such an efficient symmetry-breaking mechanism involves relative intermolecular displacements along the rubrene short axis. In the adiabatic representation, the $C_{2h}$ $\pi$-stacking corresponds to a conical intersection (zero-gap seam) \cite{CI}, and the symmetry-breaking induces an avoided-crossing between the adiabatic states (Fig. 4C).

We have thus calculated the potential energy curve and the TT-XT and TT-CT electronic couplings with respect to the displacement along the short axis (Fig. 4A and 4B), using DFT with the B3LYP functional and the Grimme's dispersion correction. From the linear dependence of the electronic coupling with displacement, one can extract the corresponding off-diagonal (non-local) electron-phonon coupling, $V_{IJ} = \lambda_{IJ} X $ ($\lambda_{TT-XT} =$ 0.002 eV, $\lambda_{TT-CT} =$ 0.003 eV), while the fit with an harmonic oscillator of the ground-state potential energy curve yields the frequency ($\sim$0.007 eV) of the inter-molecular symmetry-breaking mode, $X$. 

The quantum dynamics simulations have been performed by including in the Hamiltonian the symmetry-breaking inter-molecular mode in addition to the intra-molecular modes. Temperature effects are built in by considering populations of the initial vibrational excitations (i.e., phonon number, $N_{phonon}$) of the symmetry-breaking mode from the Bose-Einstein distribution, $ \langle N_{phonon} \rangle = 1/\{ \exp(\beta\omega)-1 \} $, where $\beta$ is the inverse temperature. Our quantum dynamics calculations indicate an increase in the SF rate with increasing temperature (Fig. 4D), consistent with the measured thermally activated behaviour \cite{RubrenePRB}. At 0 K, the vibrational wavefunction is localized around the $C_{2h}$ region where SF is inefficient, since the off-diagonal vibronic couplings ($\lambda_{TT-XT}$ and $\lambda_{TT-CT}$) are small in this system. At higher temperatures, the symmetry-breaking mode broadens the wave packet over a region where the couplings, $V_{TX}(X)$ and $V_{TC}(X)$, become large. In contrast to TIPS-pentacene, SF in rubrene is: (i) incoherent  (Fig. 3A), (ii) mediated by the direct 2-electron transition with negligible contribution from the superexchange pathway (Fig. 4E), and (iii) occurs around a conical intersection with the wave packet on TT being affected by destructive interference due to the Berry phase \cite{SM, CI}.

\begin{figure}[!t]
\includegraphics[width=8.0cm]{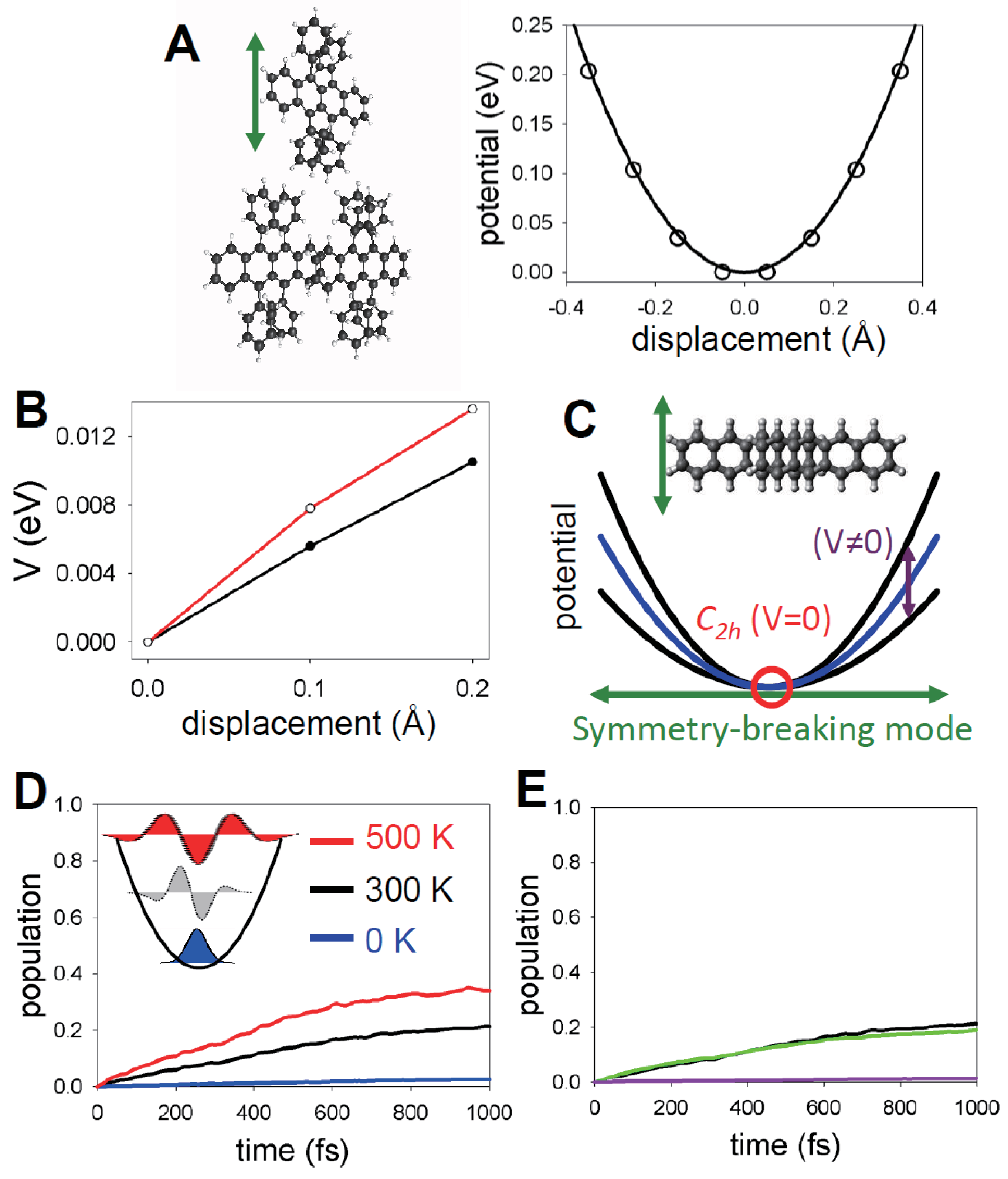}
\caption{
\label{fig4} 
(Color online)
(A) Potential curve with respect to displacement along the short molecular axis as obtained from DFT calculations of three rubrene molecules extracted from the crystal. To account for the steric hindrance from the other side of molecular layer, the mirror image of the calculated potential curve is added. (B) TT-XT (black) and TT-CT (red) electonic couplings by MRMP2. (C) Schematic illustration of the conical intersection at the $C_{2h}$ $\pi$-stacking (red circle) and the avoided crossing due to symmetry-breaking, where the black and blue parabolas illustrate the adiabatic potentials and the seam of diabatic potentials, respectively. (D) TT population during SF in rubrene considering a bright initial exciton. Two inter-molecular modes (sites 1-2, and sites 2-3) are considered, where $ \langle N_{phonon} \rangle $ are 0 (blue), 3 (black), and 6 (red). (E) The black line is identical to that in Fig. 4D, which is compared with TT evolution considering only one of the direct (green) and superexchange (purple) pathways.
}
\end{figure}

In summary, we have proposed a nonadiabatic quantum dynamical model of SF fully parametrized by first principles calculations, and successfully clarified the mechanisms underlying the ultrafast and thermally activated SF in $\pi$-stacked acenes. The present analysis provides conclusive evidence that in TIPS-pentacene both the direct and superexchange pathways mediate SF via an avoided-crossing. Notably, the coherent SF via higher-lying CT states and its dominant contribution to the ultrafast dynamics are not intuitively obvious without the explicit quantum dynamical analysis. While the slipped-stacked TIPS-pentacene results in strong electronic couplings, the electronic couplings between the TT and other states vanish at the equilibrium $C_{2h}$ stacking of rubrene. SF of rubrene turned out to be driven by thermal excitations of symmetry-breaking vibrations which enhance the electronic couplings. The marked difference in the SF mechanisms due to the packing symmetry is an interesting feature of $\pi$-stacked acenes which can inspire concepts of molecular design for controlling SF.

H.T. was supported by Fellowship for Invited Professor, FNRS, Belgium. 
Advanced Institute for Materials Research (AIMR) is supported by World Premier International Research Center Initiative (WPI), MEXT, Japan. 
D. B. is FNRS Research Fellow. 
M. H.-R. acknowledges a fellowship within the postdoctoral program of the Alexander von Humboldt foundation.
Support by the Deutsche Forschungsgemeinschaft (DFG) in the framework of the project BU-1032-2 is gratefully acknowledged.

\end{document}